# Superconducting series nanowire detector counting up to twelve photons

Zili Zhou,[1,*] Saeedeh Jahanmirinejad,[1] Francesco Mattioli,[2] Döndü Sahin,[1] Giulia Frucci,[1] Alessandro Gaggero,[2] Roberto Leoni,[2] and Andrea Fiore[1]

[1] *COBRA Research Institute, EindhovenUniversity of Technology, PO Box 513, 5600 MB Eindhoven, The Netherlands*
[2] *Istituto di Fotonica e Nanotecnologie, CNR, Via Cineto Romano 42, 00156 Roma, Italy*
[*] z.zhou@tue.nl

**Abstract:** We demonstrate a superconducting photon-number-resolving detector capable of resolving up to twelve photons at telecommunication wavelengths. It is based on a series array of twelve superconducting NbN nanowire elements, each connected in parallel with an integrated resistor. The photon-induced voltage signals from the twelve elements are summed up into a single readout pulse with a height proportional to the detected photon number. Thirteen distinct output levels corresponding to the detection of $n$=0-12 photons are observed experimentally. A detailed analysis of the excess noise shows the potential of scaling to an even larger dynamic range.

## 1. Introduction

Photon-number-resolving (PNR) detectors have attracted a large interest in the last decade. They play a key role in many fields such as linear-optics quantum computing [1] and quantum communication [2]. Ultimately, a PNR detector with a large dynamic range would represent an ideal photon detector combining single-photon sensitivity with a linear response. However, making a PNR detector that meets the requirements of these applications is challenging. It should have high efficiency, high speed, low jitter, low noise, sensitivity at telecommunication wavelengths and the ability of resolving photon numbers with a large dynamic range. None of the existing PNR detectors meet all of these standards. For instances, transition edge sensors [3], charge integrated photon detectors [4] and PNR detectors based on time-multiplexing [5] are limited by their poor timing properties, arrays of InGaAs single photon avalanche detectors (SPAD's) [6, 7] and arrays of silicon photomultipliers [8] have high dark count rates (DCR), visible light photon counters [9] are not sensitive at telecommunication wavelengths, and InGaAs SPAD's with self-differencing circuits have a limited PNR dynamic range [10]. Superconducting single photon detectors (SSPD's) [11] are well known for their leading performance in the photodetection at telecommunication wavelengths, providing high quantum efficiency (QE), short response time, low timing jitter, and low DCR [12]. PNR detectors based on SSPD's will benefit from these advantages and are promising to outpace the other existing PNR techniques.

The SSPD's operated with the conventional readout scheme (an amplifier chain with a 50 $\Omega$ input impedance) do not provide the PNR functionality. Indeed, since the resistance of the NbN nanowire after photon absorption (~$10^3$ $\Omega$) is much larger than the load resistance (50 $\Omega$), the absorption of more than one photon in the wire results in a readout pulse with nearly the same height as the one produced by the absorption of a single photon. Therefore the SSPD's give a binary response (either '0 photon' or '$\geq$ 1 photon') to the number of incident photons. Lowering the bias current of the SSPD will bring it to the multi-photon detection regimes [11, 13-15]. In this mode, the SSPD operates as a threshold detector (responding to $\geq n$ photons) but not as a PNR detector. Integrating an SSPD with a high impedance preamplifier may in principle enable the SSPD to have the PNR capability [16]. However, the PNR resolution with this method was limited due to the latching problems [17]. An alternative way is spatial multiplexing [18, 19]. In this case, a number of SSPD elements are arranged in an array and equally illuminated under the input light beam. Each element acts as a single-photon detector and their responses are read either separately [18] or together [19] in the output. The former technique, i.e. reading the responses from a number of SSPD elements separately, requires the same number of readout electronic sets. Therefore it is not scalable to large photon numbers due to the correspondingly increasing experimental complexity. In contrast, the latter approach, i.e. reading the responses from different elements summed up together as a single readout signal, can avoid this problem. In this case, the detected photon number is measured from the height of the single readout signal. As reported in Ref [19], an array of six superconducting nanowire elements connected in parallel, named parallel nanowire detector (PND), was able to resolve up to five photons with a record PNR performance. However, the PND has a drawback of current redistribution problems due to its parallel design. The bias current from the firing elements is partially redirected to the unfiring elements. This phenomenon introduces fake photon detections and consequently limits the dynamic range [20]. In order to solve this problem, Jahanmirinejad *et al.* [21] proposed a novel design of a series connection of *N* superconducting nanowires named *series nanowire detector* (SND), which solves the current redistribution problem and is in principle scalable to large photon numbers. The first experimental demonstration of the SND, as a proof of principle, was reported in [22] with four elements (namely 4-SND), capable of resolving up to four photons. A waveguide-coupled 4-SND has also been demonstrated recently [23]. However, so far, the scalability of the SND to large photon numbers has not been proved experimentally. The excess noise observed in the 4-SND's potentially represents a threat to the scalability, as it may undermine the discrimination between photon levels at large photon numbers.

In this work, we extend the PNR dynamic range of the SND to twelve photons. The SND presented here, named as 12-SND, is based on a series array of twelve superconducting NbN nanowire elements. The photon-induced voltage pulses produced by the switching elements are summed up into a single readout pulse with a height proportional to the number of firing elements and thus to the number of detected photons. Thirteen distinct output levels corresponding to the detection of $n$=0-12 photons were obtained in the measurement, representing a record dynamic range for fast PNR detectors at telecommunication wavelengths. The device quantum efficiency (DQE) and the temporal properties of the device were characterized. Photon statistics were performed on the experimental data and showed a good agreement with the theory. A detailed analysis of the linearity and of the excess noise of the 12-SND is presented, providing valuable information on the SND operation and on the potential for further scaling the dynamic range.

## 2. Device and experimental setup

A schematic diagram of the 12-SND is shown in Fig. 1(a). Twelve elements are connected in series and each of them consists of a section of superconducting nanowire, shunted with a parallel resistor $R_p$. The working principle of each element is similar to a standard SSPD [11]. Each element is biased with a current $I_B$, slightly lower than the critical current $I_C$, using a current source. When no photon arrives, the nanowire is in the superconducting state and the $I_B$ flows through the nanowire. When a photon is absorbed, the photon energy suppresses the superconductivity in the nanowire and triggers the transition to the normal state. Since the resistance of the nanowire after photon absorption is much larger than the value of $R_p$, the $I_B$ is diverted to the $R_p$ and builds a voltage pulse across it. The photo-induced voltages of different elements are summed up in the readout resistor $R_L$ (50 Ω in this work), producing a single output voltage pulse with a height proportional to the number of firing elements, and therefore to the number of detected photons.

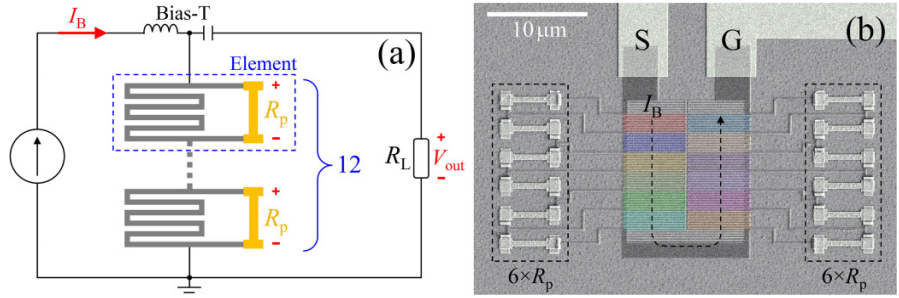

Fig. 1. (a) Schematic diagram of the 12-SND (not to scale). (b) SEM image of the 12-SND. The twelve active nanowires are highlighted in colors. The twelve $R_p$'s, the signal (S) and ground (G) contact pads, and the direction of $I_B$ are indicated. The white scale bar in the upper-left corner of the image indicates a length of 10 μm.

A scanning electron microscope (SEM) image of the 12-SND is shown in Fig. 1(b). The fabrication process of the 12-SND is similar to that of the 4-SND as reported in Ref [22]. After a 4.8 nm-thick NbN film was grown on a GaAs substrate [24] by reactive DC-magnetron sputtering, four electron-beam lithography steps were taken to fabricate the 12-SND. First, the main electrical contact pads [Ti(10 nm)/Au(60 nm)] and alignment markers were fabricated by metal evaporation and lift-off using a polymethylmethacrylate (PMMA) stencil mask. The $I_B$ flows through these contact pads [marked by S (signal) and G (ground) in Fig. 1(b)] into the device. Second, for each 12-SND, twenty-four smaller contact pads [Ti(5 nm)/Au(20 nm)] were made for the electrical contact of the twelve $R_p$'s. Third, the NbN film was patterned by reactive ion etching using hydrogen silsesquioxane as an etch mask. The twelve active NbN nanowire sections [highlighted by different colors in Fig. 1(b)] were patterned in a 12 μm × 12 μm array with a filling factor of 40%. The width of the NbN nanowires in the array is 100 nm. In the last step, twelve $R_p$'s [Ti(10 nm)/AuPd(50 nm)] were fabricated by lift-off using a PMMA stencil mask. The twelve $R_p$'s locate on the sides of the nanowire array, each of them has a design value of 50 Ω and is connected to the array through the 250 nm-wide NbN nanowires.

In the experiment, the device was kept at a temperature of 1.2 K in a VeriCold cryostat, which consists of a Pulse Tube Cooler with an additional Joule-Thompson closed cycle. According to the current-voltage (IV) characterization of the 12-SND as shown in Fig. 2, the $I_C$ of the device was 13.4 μA at 1.2 K. The typical relaxation-oscillation regime is not observed on the IV curve (solid red line) due to the presence of the twelve $R_p$'s. When $I_B$ exceeds $I_C$, the entire nanowire becomes resistive, so the measured resistance ($dV_B/dI_B$) equals to the parallel equivalent of the nanowire's normal resistance and the value of 12×$R_p$. Since the resistance of the normal nanowire is much larger than the value of $12 \times R_p$, the value of $dV_B/dI_B$ can be approximated to be $12 \times R_p$ [22, 23]. As shown in Fig. 2, by calculating the reciprocal of the slope on the IV curve, the value of $12 \times R_p$ was determined to be ~542 Ω, and thus the average value of $R_p$ was ~45.2 Ω, in good agreement with the design value of 50 Ω. The asymmetric shape of the IV curve in the normal region, which also showed hysteresis, was attributed to spurious reflections from the amplifier, which was connected to the device via the bias-T in the IV measurement and in the optical characterizations presented below.

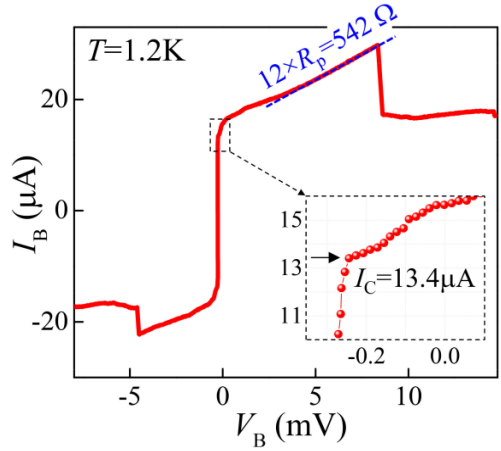

Fig. 2. IV characterization of the 12-SND at $T$ = 1.2 K with the amplifier connected to the device. The value of $12 \times R_p$ is determined by calculating the reciprocal of the slope using a linear fit (dashed blue line). The inset presents an enlarged view on the IV curve. The $I_C$ of the device was 13.4 μA.

A 1.31μm diode laser with a pulse width ($\tau_p$) of ~100 ps was used for the optical characterization in this work. The laser was triggered externally by a function generator with a repetition rate of 1 MHz. The 12-SND was illuminated by the laser through a polarization-maintaining single-mode lensed fiber mounted on a XYZ-piezo stage in the cryostat. The light spot was aligned to the center of the 12-SND and the lensed fiber tip was lifted up from its optimal focusing position to achieve a uniform illumination on the active area of the device. The full width at half maximum (FWHM) of the Gaussian spot was measured to be ~11.8 μm using the knife-edge method [25], which was performed by measuring the power of the light reflected back through the fiber tip while scanning the spot across the edge of the contact pads. Considering the size of the light spot, a ratio of ~33% of the light power from the fiber tip flowed into the 12 μm × 12 μm active area during the measurements. The optical response signal of the device was collected through the RF arm of the bias-T at room temperature, then amplified by a low-noise amplifier, and finally sent to either a 40-GHz sampling oscilloscope or a 350-MHz counter for analysis.

### 3. Optical characterizations and discussions

The optical characterization of the 12-SND was first performed with the 40-GHz sampling oscilloscope. The device was biased with an $I_B$ of 13.0 μA. The sampling oscilloscope was synchronized with the laser's trigger signal and measured the amplified output voltage signals ($V_{out}$) from the device. A low-pass filter with a cutoff wavelength of 80MHz was added in the readout circuit to improve the signal-to-noise ratio (SNR) of the output signals by removing high frequency noises. A MITEQ low-noise amplifier with 51 dB amplification, 1.1 dB noise figure, and a passband of from 0.5 to 500 MHz, was used in this measurement.

The histograms of the output signals obtained in a power range of 0-64 nW are shown in Fig. 3. They were recorded within a 50 ps time window in order to make the DCR negligible. Thirteen distinct output levels corresponding to the detections of 0-12 photons were obtained, showing a large dynamic range of the 12-SND. A detailed analysis of the data of Fig. 3 is presented in the following.

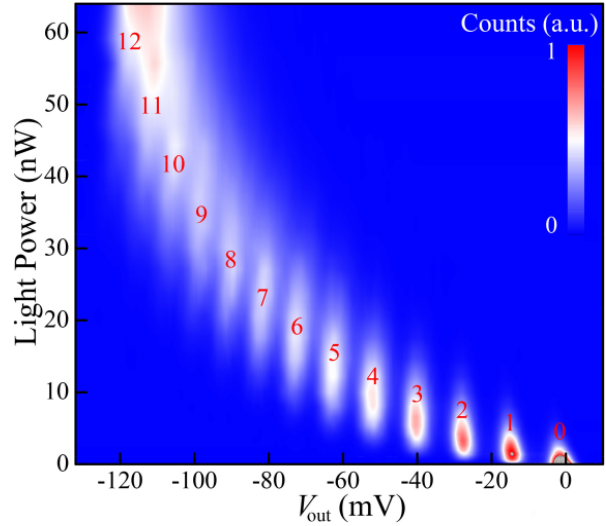

Fig. 3. The histograms of the output signals obtained in a light power range of 0-64 nW. Thirteen distinct output levels corresponding to the detections of 0-12 photons (mark by red numbers) were obtained, showing a large dynamic range of the 12-SND.

First, the distribution of the measured photon numbers was studied. The histograms at different input light powers were fitted by a sum of Gaussian peaks, using the amplitude, position and width of the peaks as fitting parameters for each power. For examples, Fig. 4(a) and 4(b) show two profiles of Fig. 3 at two different powers of 5.33 nW and 20.59 nW together with their fittings, corresponding to the detections of $n$=0-6 and $n$=3-10 photons, respectively.

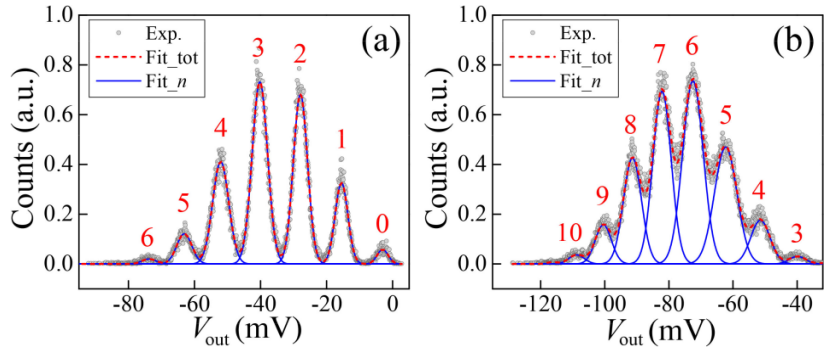

Fig. 4. Two profiles (gray dots) of Fig. 3 at two different light powers of 5.33 nW (a) and 20.59 nW (b). They are fitted using a sum of Gaussian peaks, corresponding to the detections of $n$=0-6 (a) and $n$=3-10 (b) photons, respectively. The single Gaussian fits (Fit_$n$) are shown as solid blue lines and their sum (Fit_tot) is depicted as a dashed red line.

The normalized area of each fitting Gaussian peak in the $n$th output level corresponds to the probability $P$ of detecting $n$ photons. According to Fitch *et al*. [5], for a coherent laser source uniformly incident on an array of $N$ detection elements, the probability $P$ of detecting $n$ photons can be written as,

$$P_\eta^N(n|\bar{\mu}) = \sum_{m=n}^{\infty} \frac{N!}{n!(N-n)!} \frac{(\eta\bar{\mu})^m e^{-\eta\bar{\mu}}}{m!} \times \sum_{j=0}^{n}(-1)^j \frac{n!}{j!(n-j)!}\left[1-\eta+\frac{(n-j)\eta}{N}\right]^m \quad (1)$$

Where $\eta$ is the quantum efficiency of the array and $\bar{\mu}$ is the average photon number per light pulse (proportional to the input light power). The value of $P_\eta^N(n|\bar{\mu})$ obtained at different light powers in the experiment are extracted from Fig. 3 and presented in Fig. 5 as black dots. The measured $P_\eta^N(n|\bar{\mu})$ was then fitted by calculations (red stars in Fig. 5) based on Eq. (1) using a single fitting parameter $\eta$ for each power. A good agreement between the measurements and the calculations is achieved over the entire power range.

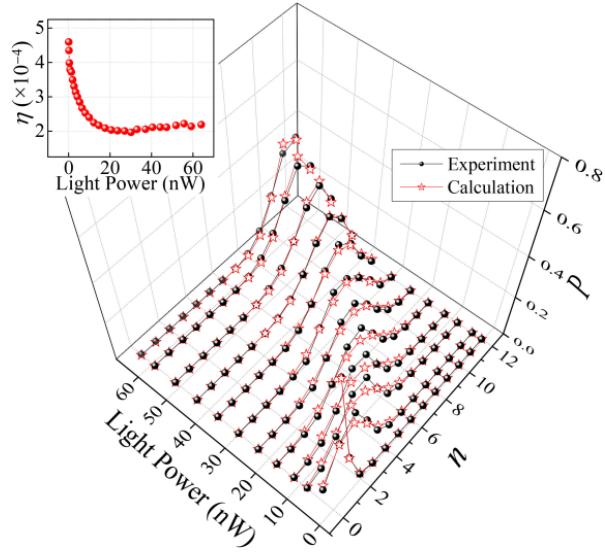

Fig. 5. The value of $P_\eta^N(n|\bar{\mu})$ (black dots) at different light powers extracted from Fig. 3 and the fitting (red stars) based on Eq. (1) using a single fitting parameter $\eta$. Inset: the value of $\eta$ obtained from the fitting is plotted as a function of the light power.

We notice that when the light power increases, $\eta$ first decreases and then maintains the minimum value at higher powers, as shown in the inset of Fig. 5. We attribute this observation to the non-uniformity of the elements' efficiency of the 12-SND. With low powers (so $\bar{\mu}$ is low), only the most efficient elements fire so we see the highest value of $\eta$ in the low power range; when $\bar{\mu}$ increases, the less efficient elements also participate in the detection so the average efficiency drops and we see a lower value of $\eta$ in the higher power range. There are three possible reasons for the non-uniformity of the element's efficiency. The first is due to the fact that the illumination is not perfectly uniform on the twelve elements. The illumination uniformity can be improved by either further increasing the Gaussian spot size or changing the spatial arrangement of the elements [26]. The second is due to the imperfections in the nanowires from the fabrication. The variation of wire's quality or geometry results in a variation of the efficiency of different elements. This can be avoided by further optimizing the fabrication process. The third reason is due to the decreasing of the bias current in the unfiring elements ($I_{uf}$) when other elements fired. As shown in Fig. 6(a), we calculated the value of $I_{uf}$ as a function of time based on the electro-thermal model reported in Ref [21] using the parameters of the present 12-SND. The value of $I_{uf}$ obviously decreases in the cases of detecting $n$=1-11 photons due to the partial redistribution of the $I_B$ to the 50 Ω load [21]. The inset provides an enlarged view of $I_{uf}$, where the temporal profile of the incident pulse ($\tau_p$=100 ps) is also indicated. Since the value of $I_{uf}$ decreases in the time window where the detection events take place, the efficiency of the unfiring elements will decrease. Although this is not desirable for PNR detection, it can be avoided by using a much shorter light pulse or using a larger load resistance (e.g. a pre-amplifier with high input impedance [21]). As an example, the calculation of $I_{uf}$ for $R_L$=1 MΩ was performed and is presented in Fig. 6(b), keeping the other parameters unchanged. In this case, the value of $I_{uf}$ remains constant so the efficiency of the unfiring elements will not decrease as a function of time.

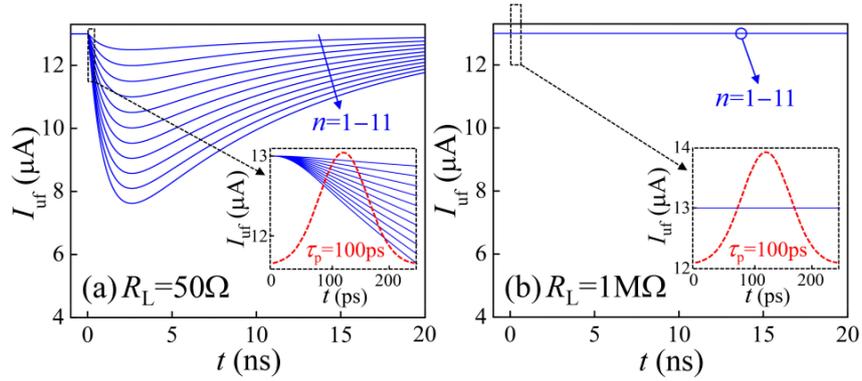

Fig. 6. The calculated value of $I_{uf}$ (solid blue lines) as a function of time in the cases of detecting $n$=1-11 photons, for $R_L$ = 50 Ω (a) and $R_L$ = 1 MΩ (b), respectively. The inset provides an enlarged view of $I_{uf}$ where the temporal profile of the incident pulse ($\tau_p$=100 ps, dashed red line) is also indicated.

We then investigated the dependence of the detection noise on the detected photon number and on the light power. The noise $V_N$ on the output levels, defined as the FWHM of the fitting Gaussian peak, is extracted from Fig. 3 and plotted in Fig. 7 for $n$=0-11 in the power range of 1.95-30.11 nW ($V_N$ of the $n$=12 peak is not shown here since the $n$=12 peak is only observed in the power range of >30.11 nW, where the fitting becomes unreliable due to the large value of $V_N$ at high powers).

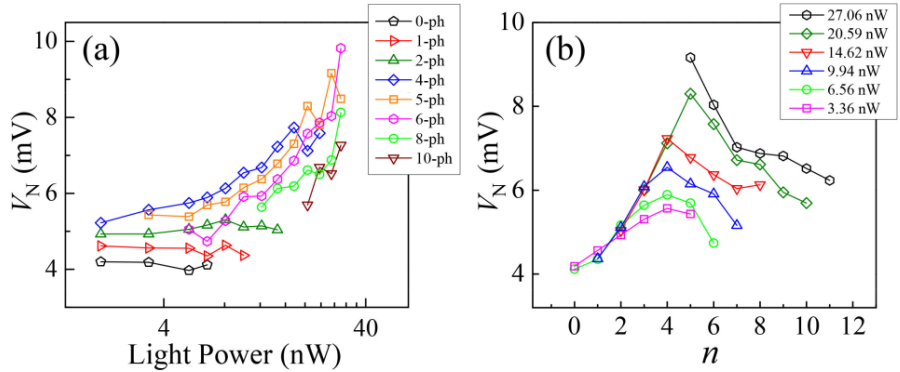

Fig. 7. (a) The value of $V_N$ is plotted as a function of the input light power for a few $n$ in the 0-10 range. (b) The value of $V_N$ is plotted as a function of $n$ for different input light powers.

On the one hand, as shown in Fig. 7(a), the value of $V_N$ for the *n*th detection increases with the light power. Indeed we have also observed that the noise of the photoresponse pulses from a standard SSPD increases with the light power. A similar observation was reported in Ref [17]. We attribute this phenomenon to the variation of the film's local characteristics or geometry (i.e. thickness and width) along the nanowires. In general, the narrower sections of the wire have higher current density and are more easily triggered by single photons. Thus, at low light powers, the output pulse height will be determined by the normal resistance of these narrower sections. When the light power increases, the wider sections of the wire will be triggered by multi-photons, contributing to the photocounts. Since the wider sections have a lower normal resistance, the distribution of the photoresponse pulse height will be broadened at higher powers. This type of noise critically depends on the uniformity of the film along the wire. It can be suppressed by optimizing the film deposition and device fabrication processes.

On the other hand, as shown in Fig. 7(b), the value of $V_N$ varies with *n* at fixed light powers, indicating excess noise in our PNR detection. Interestingly, for all powers, the value of $V_N$ first increases with *n*, reaching a maximum for $n \approx$ 4-6, depending on the power, then decreases. This dependence indicates a surprising suppression of the excess noise, which is key to the measurement of large photon numbers.

We attribute the excess noise to three possible origins. The first is due to the thermal noise of the amplifier. The thermal noise produced by the equivalent noise current generator of the amplifier depends on the detector's impedance, which dynamically varies during the detection events and depends on the number of absorbed photons. The second source of noise is due to the accumulation of the intrinsic noises from each of the *n* detection events. As discussed above, the intrinsic noise from each element is related to the inhomogeneities of the nanowires and increases with the light power. With a fixed light power, the noises in distinct wires are uncorrelated; the variance of their sum scales as *n* and their total contribution to the noise is expected to scale as $\sqrt{n}$. The third source of noise results from the variation of the height of the output pulses from different elements, which is due to the variation of $R_p$ and of the nanowire's normal resistance in different elements. As it is not feasible to directly measure the pulse height of an individual element, we simply assume a Gaussian distribution of the pulse height for the twelve elements. The number of the elements ($N_{element}$) producing a pulse height of $V_{out}^1$ in our example is shown on the right axis of Fig. 8(a) (gray bars) as a function of $V_{out}^1$. This $N_{element}(V_{out}^1)$ Gaussian distribution is assumed to have a maximum at $V_{out}^1 = 1$ and a FWHM of 0.1. Based on $N_{element}(V_{out}^1)$, we can calculate the distribution of $P_n(V_{out}^n)$, where $P_n$ is the probability of detecting a pulse with the height of $V_{out}^n$ in the *n*-photon detection event (*n*=0-12). When *n*=0, $V_{out}^0 = 0$ and the width of the distribution equals zero. When *n*=1, the distribution of $P_1(V_{out}^1)$ reproduces the distribution of $N_{element}(V_{out}^1)$. When *n*>1, each *n*-photon detection is a combination of *n* one-photon detections corresponding to any *n* of the twelve elements. The value of $P_n(V_{out}^n)$ equals to the sum of the probabilities for all the possible combinations which produce $V_{out}^n$. The calculation of $P_n(V_{out}^n)$ for *n*=0-12 has been done based on the above rules. Four examples for *n*=1, 4, 6 and 11 are plotted as histograms (gray bars) in Fig. 8(a)-(d), respectively. Each calculated $P_n(V_{out}^n)$ distribution is fitted by a Gaussian peak [red lines in Fig. 8(a)-(d)]. The FWHM of the Gaussian peak, which represents the excess noise, is plotted as a function of *n* in Fig. 8(e). Interestingly, the shape of the $P_n(V_{out}^n)$ distribution for the *n*-photon event is identical to that of the (12−*n*)-photon event. For instance, the one-photon event shown in Fig. 8(a) reproduces the distribution of the eleven-photon event shown in Fig. 8(d), despite of the shift of $V_{out}^n$. It is also interesting to see that the excess noise drops to zero when *n*=12 as shown in Fig. 8(e). Indeed, if all the twelve elements are triggered, only a single value of $V_{out}^{12}$ is possible, so no noise appears for $n$=12. The calculated $n$-dependence of excess noise qualitatively agrees with the experimental data. The calculation [Fig. 8(e)] gives the highest noise at $n$=6, while in the experiment [Fig. 7(b)] the highest noises take place in the range of $n$=4-6. The agreement indicates that the statistical distribution of $R_p$ and of the nanowire's normal resistance plays an important role in the observed excess noise. We notice that at lower powers, the excess noise tends to drop to the minimum at $n < 12$, instead of $n$=12 in the calculation. This might be due to the fact that the efficiencies of the twelve elements were not uniform so that at lower powers only the most efficient elements participated in the detection.

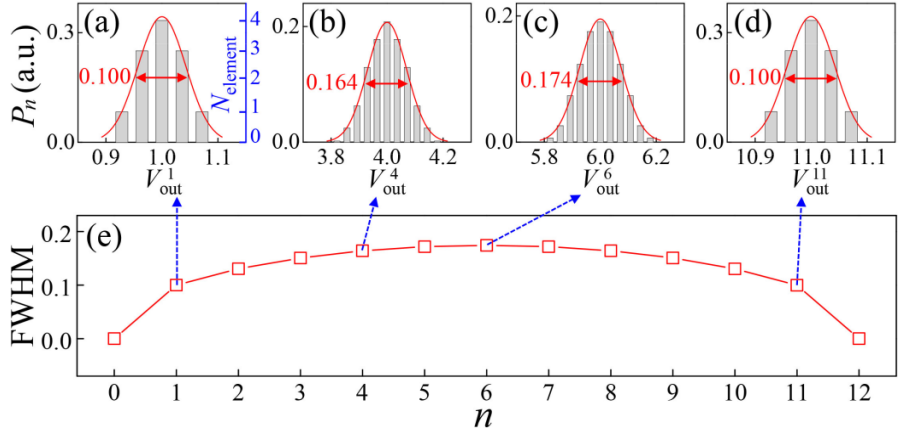

Fig. 8. The Gaussian distribution of $N_{element}(V_{out}^1)$ and the corresponding distribution of $P_1(V_{out}^1)$ for $n$=1 are shown as histogram (gray bars) in (a) to the right and the left axis, respectively. The distribution of $P_n(V_{out}^n)$ for $n$=0-12 is calculated based on $P_1(V_{out}^1)$. The examples for $n$=4, 6 and 11 are plotted as histograms (gray bars) in (b), (c) and (d), respectively. Each $P_n(V_{out}^n)$ distribution is fitted by a Gaussian peak (red lines). The FWHM of the fitting Gaussian peak, which represents the excess noise, is plotted as a function of $n$ in (e) for $n$=0-12. The FWHM for the examples of $n$=1, 4, 6 and 11 is marked in (a)-(d), respectively.

The linearity of the 12-SND's output was then investigated. The average height ($H$) of the output voltage levels was extracted from Fig. 3 and is plotted as a function of $n$ in log-log scale in Fig. 9(a). A power-law fit to the measurement, defined as $H=A\cdot n^\alpha$ ($A$ and $\alpha$ are fitting parameters), is also plotted. The obtained value of $\alpha$ is 0.84, providing a good linearity for PNR detection.

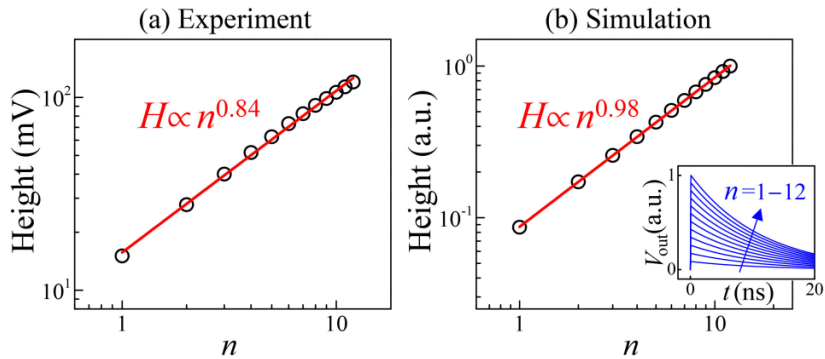

Fig. 9. (a) The value of $H$ (black dots) extracted from Fig. 3 is plotted as a function of $n$ in log-log scale. A power-law fit (red line) to the measurement, defined as $H=A\cdot n^\alpha$, is also plotted, giving $\alpha$ =

0.84. Inset of (b): the calculated $V_\text{out}$ is plotted as a function of time for $n$=1-12 using the electro-thermal model [21]. Main panel of (b): the value of $H$ (black dots) extracted from the inset is plotted as a function of $n$ together with a power-law fit (red line), giving $\alpha$ = 0.98.

We calculated the value of $V_\text{out}$ as a function of time for $n$=1-12 using the electro-thermal model [21] with the parameters of the present work, and plot the results in the inset of Fig. 9(b). The height of these pulses, i.e. the maximum value of $V_\text{out}$, is extracted and plotted in the main panel of Fig. 9(b). By fitting the calculated $H$ as a function of $n$ in log-log scale, we obtained an $\alpha$ of 0.98, showing a better linearity than in the experiment. We attribute the difference of the $\alpha$ values between the experiment and the calculation to the inhomogeneities of the nanowires, which is also one of the noise sources as discussed above. When the light power increases, the variation of nanowire's width does not only broaden the $V_\text{out}$ distribution, but also reduces the average value of $V_\text{out}$, which represents the value of $H$. This is because wider sections are triggered at higher powers and the height of pulses originated from them is smaller than that from the narrower sections.

To further confirm the PNR functionality of the 12-SND, we studied the count rate (CR) dependence on the light power, using the 350-MHz counter. As shown in Fig. 10, the CR was measured as a function of the input light power by setting different trigger levels of the counter (the DCR was subtracted from the data). The trigger level was chosen between the $n$th and the ($n$+1)th output levels in Fig. 3 so that the counter only recorded the '≥$n$-photon' responses. According to Eq. (1), if the 12-SND has the PNR functionality, the measured CR at low light powers (i.e. $\eta\bar{\mu} \ll 1$) should be approximately proportional to $(\eta\bar{\mu})^n$ and thus proportional to the $n$th-order of the light power. This is confirmed by our results as shown in Fig. 10. In the power range of up to 4 nW, which approximately corresponds to $\eta\bar{\mu} < 1$, the slope of the curves agrees with the value of 1, 2, ⋯, and 7 for the corresponding photon regimes. For higher powers (>4 nW), the approximation does not apply because the value of $\eta\bar{\mu}$ becomes larger than 1.

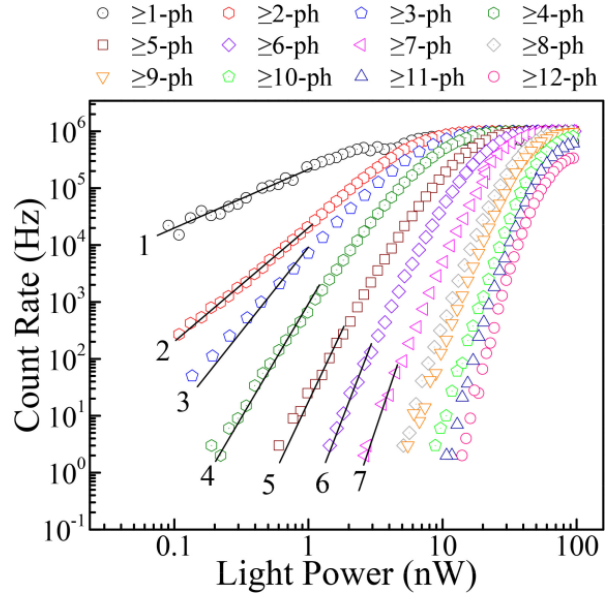

Fig. 10. The measured CR is plotted as a function of the light power by setting different trigger levels of the counter, corresponding to the detections of '≥$n$-photons' ($n$=1-12). The slopes of the curves at lower powers agree well with the $(\eta\bar{\mu})^n$ dependence (indicated by solid black lines), providing a proof for the 12-SND's PNR functionality.

Moreover, the DQE of the 12-SND was measured. The DQE is defined as the CR measured by the counter whose trigger level is set between the 0th and 1st output levels, divided by the number of photons per second incident on the active area of the device. After aligning the light's polarization direction parallel to the nanowires [27], the DQE was measured as a function of $I_B$ as shown in Fig. 11, reaching a maximum value of ~0.17% at $I_B$=13.2 μA. The low DQE of the present device is due to the low optical absorptance (in the order of few percent [24]) of NbN nanowires on GaAs substrates, and also due to the inhomogeneities of the NbN nanowires which effectively reduced the device's active area [28, 29]. The DQE can be improved either by changing the substrate [24] or by optimizing the quality of the NbN film and the fabrication process.

The temporal response of the 12-SND was also characterized. As shown in the inset of Fig. 11, the photoresponse pulse for the 12-photon detection was recorded by the 40-GHz sampling oscilloscope. A chain of two Mini-circuit low noise amplifiers, both of which had 13 dB amplification, 2.6 dB noise figure, and 0.02-3 GHz passband, was used in this measurement for a better temporal resolution. The 80 MHz low-pass filter was removed from the readout circuit in this case. As shown in the inset of Fig. 11, the dashed blue line is the calculated photoresponse curve for the 12-photon detection [also shown in the inset of Fig. 9(b)], giving a fall time $\tau_{fall}$ of ~11.3 ns, which enables a maximum repetition rate of ~30 MHz. After applying a band-pass filter (0.02-3 GHz, corresponding to the passband of the amplifiers) to the dashed blue line, the result (solid red line) presents a good agreement with the measurement. The system time jitter was measured at the leading edge of the photoresponse pulse to be ~89 ps, including the jitter of the 12-SND, of the laser and of the amplifiers. We note that with the present 12-SND, a trade-off exists between the temporal performance and the SNR in the PNR operation, depending on whether the low-pass filter is used in the circuit. However, this will not intrinsically limit the performance of the SND's. Since the noise was mainly due to the fabrication imperfections, we can increase the SNR by improving the uniformity of the device, releasing the need for a low-pass filter.

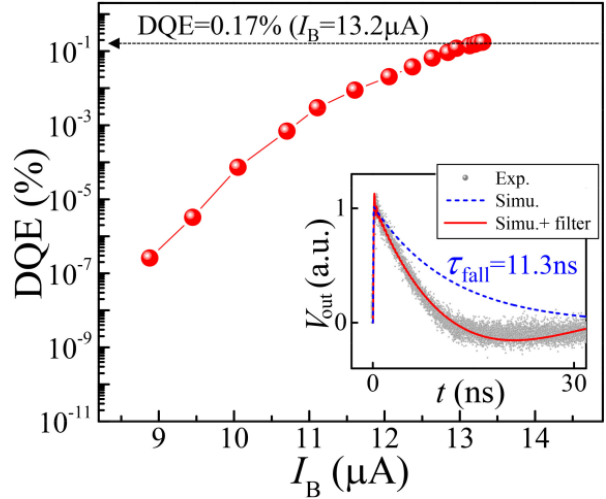

Fig. 11. The DQE is plotted as a function of $I_B$, reaching a maximum value of about 0.17% at $I_B$ = 13.2 μA. Inset: photoresponse pulse for the 12-photon detection recorded by the 40-GHz sampling oscilloscope. The dashed blue line is the calculated photoresponse pulse for the 12-photon detection event, giving a fall time $\tau_{fall}$ of ~11.3 ns. After applying a band-pass filter (0.02-3 GHz, corresponding to the passband of the amplifiers) to the simulation, the result (solid red line) shows a good agreement with the measurement.

## 4. Conclusions

In summary, we presented a PNR detector based on a series connection of twelve superconducting NbN nanowires. This detector, named as 12-SND, provides a single output pulse with a height proportional to detected photon numbers. The 12-SND is able to resolve up to twelve photons, providing a record dynamic range among the existing fast telecom-wavelength PNR detectors. The 1/e decay time of the device was ~11.3 ns, enabling a repetition rate of ~30 MHz. The system jitter was measured to be ~89 ps. The correct PNR operation was verified by studying the count rate and the distribution of the output levels as a function of the input light power. The output voltage shows a good linearity as a function of the photon number, demonstrating the potential of this device as a linear detector. The noise analysis reveals a surprising suppression of the excess noise at high photon numbers, which is attributed to the statistical distribution of the parallel resistances and of the nanowire's normal resistance. This indicates an excellent potential of the SND of scaling to even higher photon numbers, especially if the uniformity of the nanowire and of the parallel resistances is improved. Moreover, the combination of the SND with a cryogenic preamplifier with high input impedance will further improve the performance of the SND's [21]. By using the preamplifier, the bias current in the unfiring elements will not decrease, resulting in higher output signals, improved discrimination of the voltage levels, better linearity, and higher speed.

**Acknowledgments**

The authors would like to thank R. Gaudio for taking the SEM image, P. A. M. Nouwens, M. van Vlokhoven, F. W. M. van Otten, and T. Xia for technical support. This work was financially supported by NanoNextNL, a micro- and nano-technology program of the Dutch Ministry of Economic Affairs and Agriculture and Innovation and 130 partners, by the European Commission through FP7 project Q-ESSENCE (Contract No. 248095) and by the Dutch Technology Foundation STW, applied science division of NWO, the Technology Program of the Ministry of Economic Affairs under project No. 10380.